\documentclass[pdflatex,sn-mathphys-num]{sn-jnl}

\usepackage{graphicx}%
\usepackage{multirow}%
\usepackage{amsmath,amssymb,amsfonts}%
\usepackage{amsthm}%
\usepackage{mathrsfs}%
\usepackage[title]{appendix}%
\usepackage{xcolor}%
\usepackage{textcomp}%
\usepackage{manyfoot}%
\usepackage{booktabs}%
\usepackage{algpseudocode}%
\usepackage[linesnumbered, ruled, vlined]{algorithm2e}
\usepackage{listings}%
\usepackage{listings}
\usepackage{color}
\usepackage{xcolor}
\usepackage{graphicx}
\usepackage{float}
\usepackage{physics}
\usepackage{comment}
\usepackage{tcolorbox}
\usepackage{rotating} 
\usepackage{bbm} 
\usepackage{upgreek} 

\usepackage{booktabs}
\usepackage[table]{xcolor}
\usepackage{geometry}
\definecolor{headerblue}{RGB}{220,230,245}
\definecolor{rowgray}{RGB}{245,245,245}

\definecolor{dkblue}{rgb}{0,0.39,0}
\definecolor{gray}{rgb}{0.66,0.66,0.66}
\definecolor{mauve}{rgb}{0.91,0.33,0.50}
\definecolor{gold}{rgb}{1,0.84,0}

\begin{document}

\title{PHANTOM: Progressive High-fidelity Adversarial Network for Threat Object Modeling}
\author[1,2]{\fnm{Jamal} \sur{Al-Karaki}}\email{Jamal.Al-Karaki@zu.ac.ae}

\author*[1]{ \fnm{Muhammad Al-Zafar} \sur{Khan}}\email{Muhammad.Khan@zu.ac.ae}

\author[1]{\fnm{Rand} \sur{Derar Mohammad Al Athamneh}}

\affil*[1]{\orgname{College of Interdisciplinary Studies}, Zayed University, \state{Abu Dhabi}, \country{UAE}}

\affil[2]{\orgname{College of Engineering}, The Hashemite University \state{Zarqa}, \country{Jordan}}

\abstract{The scarcity of high-quality cyberattack datasets poses a fundamental challenge to developing robust machine learning-based intrusion detection systems. Real-world attack data is difficult to obtain due to privacy regulations, organizational reluctance to share breach information, and the rapidly evolving threat landscape. This paper introduces PHANTOM (Progressive High-fidelity Adversarial Network for Threat Object Modeling), a novel multi-task adversarial variational framework specifically designed for generating synthetic cyberattack datasets. PHANTOM addresses the unique challenges of cybersecurity data through three key innovations: Progressive training that captures attack patterns at multiple resolutions, dual-path learning that combines VAE stability with GAN fidelity, and domain-specific feature matching that preserves temporal causality and behavioral semantics. We implement a Multi-Task Adversarial VAE with Progressive Feature Matching (MAV-PFM) architecture that incorporates specialized loss functions for reconstruction, adversarial training, feature preservation, classification accuracy, and cyber-specific constraints. Experimental validation on a realistic synthetic dataset of $100\;000$ network traffic samples across five attack categories demonstrates that PHANTOM achieves 98\% weighted accuracy when used to train intrusion detection models tested on real attack samples. Statistical analyses, including kernel density estimation, nearest neighbor distance distributions, and $t$-SNE visualizations, confirm that generated attacks preserve the distributional properties, diversity, and class separability of authentic cyberattack patterns. However, results also reveal limitations in generating rare attack types, highlighting the need for specialized handling of severely imbalanced classes. This work advances the state-of-the-art in synthetic cybersecurity data generation, providing a foundation for training more robust threat detection systems while maintaining privacy and security.}


\maketitle

\section{Introduction}\label{introduction}
The exponential growth of cyber threats in recent years has created an urgent demand for robust cybersecurity systems capable of detecting and mitigating sophisticated attacks \cite{jang2014survey,jimmy2021emerging,admass2024cyber}. Machine Learning (ML) and Deep Learning (DL) models have emerged as powerful tools for threat detection \cite{shaukat2020cyber, alzaabi2024review}, enabling automated analysis of network traffic \cite{pacheco2018towards}, system logs \cite{du2017deeplog}, and user behavior patterns \cite{alshehri2023cyberattack}. However, the effectiveness of these models hinges critically on the availability of diverse, representative training data that captures the full spectrum of attack vectors and techniques employed by adversaries.

Despite this need, obtaining high-quality cyberattack datasets remains one of the most significant challenges in cybersecurity research and practice. Real-world attack data is inherently scarce due to several factors: 
\begin{enumerate}
\item Organizations are often reluctant to share sensitive breach information due to legal and reputational concerns \cite{wu2025trust}. 
\item Privacy regulations restrict the dissemination of network traffic containing potentially identifiable information \cite{liu2015identifying}.
\item The rapidly evolving threat landscape means that historical datasets quickly become obsolete \cite{kedys2025fast}.
\end{enumerate} 
Additionally, even when attack data is available, it often suffers from severe class imbalance, with benign traffic vastly outnumbering malicious samples, leading to biased models that struggle to detect novel or rare attack patterns.

Synthetic data generation has emerged as a promising solution to address these limitations \cite{agrawal2024review,kumar2023synthetic}. By artificially creating realistic cyberattack samples, researchers can augment existing datasets, balance class distributions, and generate examples of rare or emerging threats that may not yet exist in operational environments. However, traditional synthetic data generation techniques, such as rule-based simulation and simple statistical sampling, often produce oversimplified attack patterns that lack the complexity and variability of real-world threats. Models trained on such synthetic data frequently exhibit poor generalization when deployed in production environments, as they fail to capture the nuanced behavioral characteristics of actual attackers.

Recent advances in generative modeling, particularly Generative Adversarial Networks (GANs) and Variational Autoencoders (VAEs), offer a paradigm shift in synthetic data generation. These deep generative models learn the underlying probability distribution of real data and can generate novel samples that preserve the statistical properties and complex patterns of the original dataset. GANs, through their adversarial training mechanism between a generator and a discriminator network, have demonstrated remarkable success in generating high-fidelity synthetic data across various domains, including image synthesis, natural language processing, and time-series forecasting. Similarly, VAEs utilize probabilistic latent representations to facilitate the controlled generation of diverse samples while preserving the interpretability of the learned feature space.

In this paper, we propose specialized GAN and VAE architectures tailored specifically for generating high-fidelity synthetic cyberattack datasets. Our approach addresses the unique challenges of cybersecurity data, including temporal dependencies in attack sequences, multi-modal feature distributions spanning categorical and continuous variables, and the need to preserve attack semantics while introducing realistic variations. We develop novel architectural components and training strategies that enhance the diversity, realism, and utility of generated attack samples for downstream security applications. 

This work is divided as follows:

In Sec. \ref{related work}, we describe analogous attempts to address the major challenge that we set out to address in our research.

In Sec. \ref{the proposed approach}, we described our proposed PHANTOM framework and the mechanics of how the algorithm works. 

In Sec. \ref{experiment}, we describe the experiment performed, first by describing the dataset used, second by explaining the choice and motivation for the hyperparameter values selected in the algorithm implementation, and finally by presenting the results obtained through experimentation. 

In Sec. \ref{conclusion}, we reflect upon what was achieved in this work, the drawbacks, and provide direction for future works. 

\section{Related Work}\label{related work}
In \cite{le2020generating}, the authors address the key problem for critical space systems: The lack of high-fidelity, shareable datasets that include both nominal and malicious activity. Specifically, they propose a GAN-based system that creates realistic synthetic cyberattack datasets by training on small samples of real-world nominal and malicious data and then using the generator to produce new, high-fidelity synthetic samples. They evaluate the realism of the generated data and test its usefulness across three datasets.

In \cite{rao2024adaptive}, the authors focus on improving cybersecurity in Internet of Things (IoT) and Wireless Sensor Networks (WSNs) by using GANs. Due to the rise of sophisticated threats, especially DDoS and spoofing attacks, traditional security systems are no longer sufficient. To address this, the authors propose a new GAN-based model, called \textit{Dynamic Adaptive Threat Simulation GAN} (DATS-GAN), which generates realistic synthetic cyberattack scenarios that mimic real-world attacks, thereby enabling security systems to better detect, learn from, and adapt to evolving threats. The novelty in this work lies not only in its focus on generating such datasets but also in its ability to dynamically detect cybersecurity attacks. 

In \cite{gondhi2024wgan}, the authors address cybersecurity challenges in modern power systems, particularly the threat of stealthy false data injection (FDI) attacks that can cause operational problems such as congestion and voltage instability by proposing a defense framework that uses Wasserstein Generative Adversarial Networks (WGANs) to generate synthetic Phasor Measurement Unit (PMU) data. The workflow creates uncertainty, making it more difficult for attackers to understand, predict, or exploit the system. This work is innovative because it strategically injects realistic synthetic data into the communication stream.

\section{The Proposed Approach}\label{the proposed approach}
The generation of high-fidelity synthetic cyberattack data presents unique challenges that surpass those of conventional image or text synthesis. Cyberattack patterns exhibit complex temporal dependencies, causal relationships between attack stages, multi-scale features (from packet-level to campaign-level), and highly imbalanced class distributions. To address these challenges holistically, below we introduce PHANTOM (\textbf{P}rogressive \textbf{H}igh-fidelity \textbf{A}dversarial \textbf{N}etwork for \textbf{T}hreat \textbf{O}bject \textbf{M}odeling), a multi-task adversarial variational framework specifically designed for synthesizing cyberattack data.

Our approach is predicated on three fundamental insights about cyberattack data generation:
\begin{enumerate}
\item Cyberattacks are hierarchical and manifest at multiple resolutions simultaneously, from low-level network packet features to high-level behavioral patterns. 
\item Attack semantics are causal, which implies that actions follow logical sequences that must be preserved in synthetic data to maintain realism and utility.
\item Fidelity must be multi-dimensional. This translates to temporal, behavioral, and structural aspects that must all be preserved for synthetic data to be operationally useful.
\end{enumerate}

PHANTOM addresses these insights through an integrated architecture that combines the stability of VAEs with the high-fidelity generation capabilities of GANs, incorporating domain-specific feature preservation mechanisms. At its core, PHANTOM implements a Multi-Task Adversarial VAE with Progressive Feature Matching (MAV-PFM), which operates through three synergistic components:
\begin{enumerate}
\item Unlike conventional GANs that operate at fixed resolutions, PHANTOM employs a progressive training strategy that begins with coarse-grained attack features and gradually incorporates finer-grained details. This hierarchical approach mirrors how security analysts investigate incidents—from broad indicators to specific artifacts—and ensures that both macro- and micro-patterns are faithfully reproduced.
\item The VAE component provides stable reconstruction and meaningful latent representations, while the GAN component ensures high perceptual fidelity. Crucially, both pathways share the same generator, enabling knowledge transfer between reconstruction and pure generation tasks. This dual-path approach mitigates mode collapse, which is a critical failure mode in cybersecurity contexts where rare attack types must still be generated.
\item We introduce specialized feature extractors that encode domain-specific invariants, including temporal causality, attack graph structures, and behavioral sequences. These extractors inform a novel feature matching loss that ensures synthetic attacks maintain the essential characteristics of their real counterparts, not just statistical similarity but operational realism.
\end{enumerate}

\begin{algorithm}[H]
\caption{PHANTOM}
\label{algo:phantom}
\SetAlgoLined
\DontPrintSemicolon
\textbf{input:} \;
\hspace{\algorithmicindent} real-world cyberattack dataset $\mathcal{D}=\left\{x_{i},y_{i}\right\}$ \;
\hspace{\algorithmicindent} latent dimension $Z$\;
\hspace{\algorithmicindent} batch size $m$ \;
\hspace{\algorithmicindent} progressive levels $L$ \;
\hspace{\algorithmicindent} feature extractors $\mathcal{F}=\left\{F_{\text{network}},F_{\text{temporal}},F_{\text{behavioral}}\right\}$ \;
\textbf{initialize:} $G,D,E,C$ with weights $\theta_{G},\theta_{D},\theta_{E},\theta_{C}$, replay buffer $\mathcal{B}$\;
\For{\text{current\_level} $l=1:L$}{
$\alpha_{l}\gets\text{fade\_in\_factor}(l)$ \;
$\mathcal{D}_{l}\gets\text{resize\_samples}()$\;
\For{\text{iteration} t }{
sample batch: $\left\{x_{r},y_{r}\right\}\sim\mathcal{D}_{l},z\sim\mathcal{N}(0,I),\epsilon\sim\mathcal{N}(0,\sigma^{2}I)$\;
encode: $\mu,\sigma=E(x_{r}), z_{c}=\mu+\sigma\odot\epsilon$\;
\texttt{// generate}\;
\hspace{\algorithmicindent} $x_{\text{recon}}=G(z_{c},y_{r},l,\alpha)$\Comment{reconstructed}\;
\hspace{\algorithmicindent} $y_{s}=p(y),x_{\text{syn}}G(z,y_{s},l,\alpha)$ \Comment{synthesized}\;
\texttt{// extract features}\;
\hspace{\algorithmicindent} $F_{r},F_{\text{recon}},F_{\text{syn}}$ using $\mathcal{F}$\;
\texttt{// compute losses}\; 
\hspace{\algorithmicindent} $\mathcal{L}_{\text{recon}}=||x_{r}-x_{\text{recon}}||^{2}+\beta\;\text{KL}(q||p)$ \Comment{VAE}\;
\hspace{\algorithmicindent} $\mathcal{L}_{\text{adv}}^{G}=-\mathbb{E}[D(x_{\text{syn}},x_{s})]$ \Comment{generator}\;
\hspace{\algorithmicindent}$\mathcal{L}_{\text{adv}}^{D}=\mathbb{E}[(D_{\text{syn}})]-\mathbb{E}[D(x_{r})]+\lambda_{\text{gp}}\mathcal{R}_{\text{gp}}$\Comment{discriminator}\;
\hspace{\algorithmicindent}$\mathcal{L}_{\text{fm}}=\sum_{i}\omega_{i}||\mathcal{F}_{r}^{\left(i\right)}-\mathcal{F}_{\text{syn}}^{\left(i\right)}||$\Comment{feature matching}\;
\hspace{\algorithmicindent}$\mathcal{L}_{\text{class}}=\text{CE}[C(x_{\text{syn}},y_{s})]+\text{CE}[C(x_{r}),y_{r}]$ \Comment{classification}\;
\hspace{\algorithmicindent}$\mathcal{L}_{\text{cyber}}=\mathcal{L}_{\text{temporal}}+\mathcal{L}_{\text{causal}}+\mathcal{L}_{\text{div}}$ \Comment{cyber-specific loss}\;
\texttt{// updates} \;
\hspace{\algorithmicindent} $G,E\gets\nabla(\lambda_{1}\mathcal{L}_{\text{adv}}^{G}+\lambda_{2}\mathcal{L}_{\text{recon}}+\lambda_{3}\mathcal{L}_{\text{fm}}+\lambda_{4}\mathcal{L}_{\text{class}}+\lambda_{5}\mathcal{L}_{\text{cyber}})$\;
\hspace{\algorithmicindent}$D\gets\nabla\mathcal{L}_{\text{adv}}^{D}$\;
\hspace{\algorithmicindent} $C\gets\nabla\mathcal{L}_{\text{class}}$\;
update $\mathcal{B}$ with $x_{\text{syn}}$\;
}
\texttt{// stabilization}\;
freeze $D$, refine $G$ and $E$ with $||x_{r}-G[E(x_{r})]||_{1}$
}\;
\Return Generator $G$, discriminator $D$, encoder $E$, classifier $C$
\end{algorithm}
At a high level, Algorithm \ref{algo:phantom} operates by training progressively across multiple resolution levels, starting with coarse attack features, such as packet headers, and gradually incorporating finer details, including behavioral patterns. At each level, the algorithm processes batches of real cyberattack data $\mathcal{D}_l$, which contain attack samples $x_r$ along with their corresponding labels $y_r$. The encoder $E$ compresses real attacks into latent distributions $(\mu, \sigma)$, enabling reconstruction via the generator $G$. Simultaneously, $G$ synthesizes new attacks from random noise $\mathbf{z}$ conditioned on attack parameters $\mathbf{y}s$. The discriminator $D$ distinguishes real from synthetic samples, while the classifier $C$ ensures generated attacks match their intended categories. Three specialized feature extractors $(F_{\text{network}}, F_{\text{temporal}}, F_{\text{behavior}})$ capture network, temporal, and behavioral characteristics for domain-specific feature matching.

The outer progressive loop (lines 11-34) implements hierarchical multi-resolution training, starting with coarse network features and gradually introducing finer behavioral details through the $\alpha$ parameter. At each level, batches are sampled and processed through dual generation paths: The VAE path encodes real attacks into latent distributions ($\mu,\sigma$) for reconstruction, while the GAN path synthesizes novel attacks from random noise conditioned on attack parameters $y_s$. This dual approach ensures both stable learning through reconstruction and high-fidelity generation through adversarial training. The feature extraction block applies domain-specific transforms to capture network topology, temporal patterns, and behavioral sequences—critical for maintaining cyberattack semantics.

Five specialized loss functions collectively optimize different aspects of cyberattack synthesis. The VAE loss $\mathcal{L}_\text{recon}$ ensures latent space structure and reconstruction fidelity, while $\mathcal{L}_\text{adv}$ implements Wasserstein adversarial training with gradient penalty for stable GAN dynamics. Crucially, $\mathcal{L}_\text{fm}$ preserves domain-specific characteristics by matching features across real and synthetic data in the network, temporal, and behavioral subspaces. $\mathcal{L}_\text{class}$ maintains attack type accuracy, while $\mathcal{L}_\text{cyber}$ enforces cyber-specific constraints, including temporal consistency across attack stages, causal relationships between attack actions, and diversity in generated threats. The multi-task update balances these objectives through $\lambda$ weights, while the replay buffer prevents discriminator overfitting. Finally, the stabilization phase (line 33) refines the generator-encoder pair without adversarial pressure, ensuring convergence at each resolution level before progression.

\begin{figure}[h]
    \centering
    \includegraphics[width=0.7\linewidth]{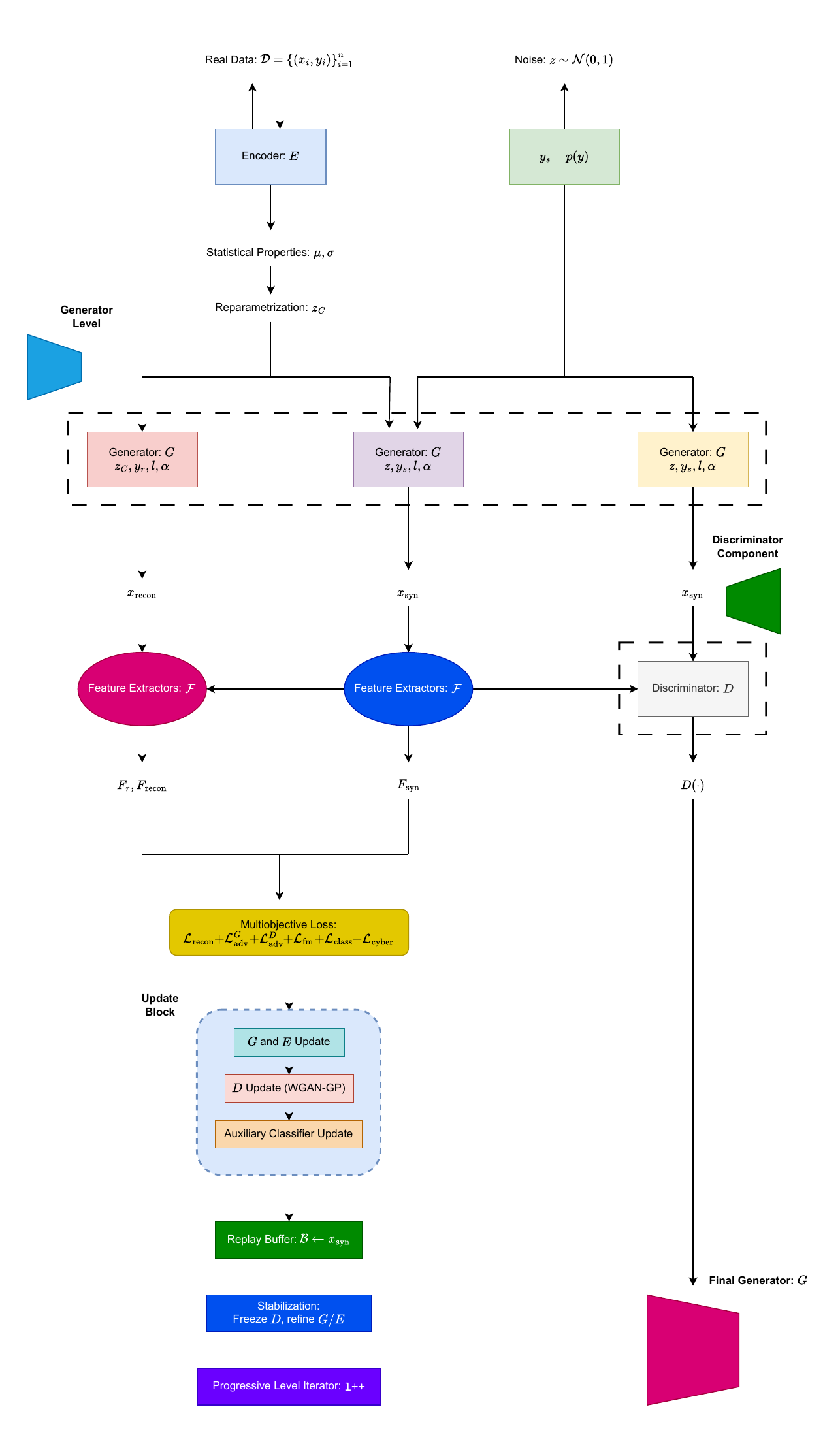}
    \caption{Network architecture diagram of the PHANTOM (\ref{algo:phantom}) algorithm.}
    \label{fig:architecture diagram}
\end{figure}

In Fig. \ref{fig:architecture diagram}, we see a graphical rendition of Algorithm \ref{algo:phantom}. The core system features parallel data flows: A VAE reconstruction path, where real attacks are encoded into latent distributions and reconstructed to ensure stability, and a GAN generation path, where random noise is transformed into novel synthetic attacks. These flows converge through a shared conditional generator that preserves attack semantics, while domain-specific feature extractors (network, temporal, behavioral) enforce cyberattack invariants through feature matching losses. The architecture is governed by a multi-objective loss function combining reconstruction fidelity (MSE + KL divergence), adversarial competition (WGAN-GP), attack classification accuracy, and cyber-specific constraints (temporal consistency, causal relationships, diversity), all orchestrated through progressive multi-resolution training that gradually refines attack patterns from coarse to fine-grained features, enabling the generation of diverse, realistic cyberattacks while maintaining the statistical and operational properties of real threat data.

We observe the spatial complexity of Algorithm \ref{algo:phantom} to be $\mathcal{O}(m\cdot(|x|+Z)+P)$, where $m$ is the batch size, $|x|$ is the input dimension, $Z$ is the latent space dimension, and $P$ is the total number of parameters across $G,D,E,C,\mathcal{F}$. Similarly, we observe the temporal complexity to be $\mathcal{O}(T\cdot L\cdot m\cdot(|x|^{2}+Z^{2}))$. We see that the spatial complexity grows linearly, $\mathcal{O}(N)$, with model capacity and batch processing requirements, making large-scale cyberattack synthesis memory-intensive but manageable with modern GPU architectures. The temporal complexity exhibits a quadratic dependence, $\mathcal{O}(N^{2})$, on feature dimensions due to attention mechanisms in cyber-specific extractors; however, progressive training mitigates this by gradually increasing resolution across levels. 

\section{Experiments}\label{experiment}

\subsection{``Real'' Dataset}
To successfully test Algorithm \ref{algo:phantom}, we generated a synthetic cyberattack dataset using this code, consisting of $100\;000$ network traffic samples across five distinct attack categories, each characterized by 40 engineered features. The dataset exhibits a deliberate realistic class imbalance mirroring real-world network environments: 70\% benign traffic ($70\;000$ samples), 15\% Denial-of-Service (DoS) attacks ($15\;000$ samples), 10\% probing activities ($10\;000$ samples), 4\% remote-to-local attacks ($4\;000$ samples), and only 1\% user-to-root privilege escalation attempts ($1\;000$ samples). Each class contains statistically distinct feature patterns derived from domain knowledge -- for instance, DoS attacks show exceptionally high source byte volumes and connection counts, while U2R attacks demonstrate prolonged durations and specific protocol usage. The features include transformed network metrics (log-scaled byte counts and normalized rates), categorical encodings of protocol types and service flags, and engineered attributes such as failed login counts and session continuity measures, providing a comprehensive representation of attack signatures.

To substantiate why we generated the dataset synthetically, it is worth noting that real-world, labeled cyberattack data of this scale and diversity is exceptionally difficult to obtain due to multiple constraints. Firstly, organizations that experience attacks rarely disclose detailed network logs due to security policies, regulatory concerns (such as GDPR \cite{voigt2017eu} and HIPAA \cite{ness2007influence}), and reputational risks. Secondly, even when incident data is shared through threat intelligence platforms, it is typically anonymized, incomplete, or lacks ground-truth labels -- security analysts often cannot definitively categorize every attack, especially novel or sophisticated threats. Thirdly, the extreme class imbalance observed here (U2R attacks constituting only 1\% of the samples) reflects reality but creates data scarcity for training robust ML models; collecting sufficient samples of rare attacks would require monitoring thousands of networks over years. Finally, operational networks cannot ethically be attacked for research purposes, making controlled experimentation with real attacks impossible.

It is worth mentioning that the synthetic generation approach enables reproducible cybersecurity research while addressing critical gaps in available data. By programmatically creating attacks with known ground truth, one can validate detection algorithms without violating privacy or raising legal concerns. The controlled class distribution enables a systematic investigation of imbalance-handling techniques, while feature engineering incorporates domain expertise on attack signatures. Importantly, it is our intention that this dataset serves as a benchmark for evaluating synthetic data generation methods like PHANTOM (Algorithm \ref{algo:phantom}) -- if a GAN can reproduce the statistical properties and class separability of this known distribution, it demonstrates the capability to generate useful synthetic data where real data is unavailable. The inclusion of realistic noise, protocol distributions, and attack-specific patterns creates a challenging testbed that bridges the gap between academic research and operational security needs, which we believe enables advancement in intrusion detection without compromising real network security or privacy.

\subsection{Hyperparameter Values}
In Tab. \ref{tab:hyperparameters}, we describe all the hyperparameters used for testing Algorithm \ref{algo:phantom} and provide a rationale for their choice. 

\begin{table}[h]
\centering
\caption{PHANTOM Algorithm Hyperparameters and Their Rationale}
\label{tab:hyperparameters}
\begin{tabular}{|p{3.5cm}|p{2.2cm}|p{7cm}|}
\hline
\textbf{Parameter} & \textbf{Value} & \textbf{Reason for Choice} \\
\hline
Latent Dimension  & $Z=64$ & Balances expressiveness (capturing complex attack patterns) and computational efficiency. Common choice in VAE/GAN literature for tabular data. \\
\hline
Batch Size & $m=64$ & Provides stable gradient estimates while fitting within GPU memory constraints. Powers of 2 optimize memory alignment on GPUs. \\
\hline
Progressive Levels & $L=1$ &Simplified for initial testing; full implementation utilizes 3-4 levels for hierarchical feature learning (packet$\to$flow $\to$session patterns). \\
\hline
Iterations per Level & $i_{\max}=500$ & Reduced for demonstration; typical training requires $5\;000$-$10\;000$ iterations per level for convergence. \\
\hline
KL Weight & $\beta=1.0$ & Standard $\beta$-VAE setting balancing reconstruction fidelity and latent space regularization for disentangled representations. \\
\hline
Gradient Penalty Weight & $\lambda_{\text{gp}}=10.0$ & Standard WGAN-GP value ensuring Lipschitz continuity of discriminator for stable adversarial training. \\
\hline
Adversarial Weight & $\lambda_{1}=1.0$ & Base weight for generator's adversarial loss relative to other objectives. \\
\hline
Reconstruction Weight & $\lambda_{2}=10.0$ & Prioritizes VAE reconstruction to ensure synthetic samples preserve essential attack characteristics. \\
\hline
Feature Matching Weight & $\lambda_{3}=5.0$ & Emphasizes preservation of domain-specific features (network, temporal, behavioral) crucial for cyberattack realism. \\
\hline
Classification Weight & $\lambda_{4}=1.0$ & Ensures generated attacks are classifiable with correct labels, maintaining attack type integrity. \\
\hline
Cyber Loss Weight & $\lambda_{5}=0.1$ & Lower weight for domain-specific losses (temporal consistency, causality) during initial training phases. \\
\hline
Learning Rate & $\eta=0.0002$ & Standard GAN learning rate from DCGAN/WGAN literature, providing stable convergence without oscillations. \\
\hline
Discriminator Beta1 & $\beta_{1}^{D}=0.0$ & WGAN-GP recommendation (first momentum coefficient) for discriminator to prevent mode-seeking behavior. \\
\hline
Generator/Encoder Beta-1 & $\beta_{1}^{G},\beta_{1}^{E}=0.0$ & Consistent with WGAN-GP architecture for stable generator training against critic. \\
\hline
Classifier Beta-1 & $\beta_{2}^{C}=0.5$ & Standard Adam setting for auxiliary classifier to balance exploration and exploitation. \\
\hline
Beta-2 (all models) & $\beta_{2}=0.9$ & Standard second momentum coefficient for Adam optimizer across all components. \\
\hline
Feature Extractor Dimension & $|\mathcal{F}|=32$ & Dimensionality for domain-specific feature representations, balancing information retention and model complexity. \\
\hline
Feature Matching Weights & $\footnotesize{\omega_{i}=\left[1.0, 1.0, 1.0\right]}$ & Equal importance for network, temporal, and behavioral feature preservation in initial implementation. \\
\hline
Label Prior & Uniform: $\mathcal{U}$ & Assumes balanced sampling across attack types; in practice, would follow empirical distribution from training data. \\
\hline
Fade-in Factor & $\alpha=l/L$ & Linear progression from coarse to fine features in progressive training paradigm. \\
\hline
Noise Scale & $\epsilon\sim\mathcal{N}(0,1)$ & Standard Gaussian noise for VAE reparameterization trick and latent space sampling. \\
\hline
\end{tabular}
\end{table}

\subsection{Results}
The classification report in Tab. \ref{tab: results table} presents the performance of a trained intrusion detection model when evaluated on a real cyberattack test set after being trained on PHANTOM-generated synthetic data. The results demonstrate strong overall performance with a 98\% weighted accuracy and excellent F1-scores ($1.00$) for the majority classes (Classes 0 and 1), indicating that the synthetic data successfully preserves the distinctive patterns of common attack types such as DDoS and malware. However, the complete failure on Class 4 ($\text{precision}=0.00$, $\text{recall}=0.00$, $\text{F}1=0.00$) reveals a critical limitation: PHANTOM failed to generate representative samples for rare attack types, likely due to insufficient examples in the training distribution. The disparity between the high weighted average (98\%) and lower macro average (77\%) highlights the class imbalance problem common in cybersecurity, where performance metrics weighted by class prevalence can mask poor detection of minority attack classes. This finding highlights the necessity for specialized techniques in synthetic data generation to ensure adequate representation of rare yet critical threats such as advanced persistent threats (APTs).

\begin{table}[h!]
\centering
\caption{Classification report -- Synthetic data vs. real test set. $TP=$ True Positive, $FP=$ False Positive, $FN=$ False Negative.}
\setlength{\tabcolsep}{12pt}
\renewcommand{\arraystretch}{1.3}
\rowcolors{2}{rowgray}{white}
\begin{tabular}{lcccc}
\rowcolor{headerblue}
\toprule
\textbf{Class} & \textbf{Precision} & \textbf{Recall} & \textbf{F1-Score} & \textbf{Support} \\
\rowcolor{headerblue}
 & $\left(\frac{TP}{TP+FP}\right)$ & $\left(\frac{TP}{TP+FN}\right)$ & $\left(\frac{2\cdot TP}{2\cdot TP+FP+FN}\right)$ & $(TP+FN)$ \\
\toprule
0 & 1.00 & 1.00 & 1.00 & $14\;000$ \\
1 & 1.00 & 1.00 & 1.00 & $3\;000$ \\
2 & 0.88 & 0.99 & 0.93 & $2\;000$ \\
3 & 1.00 & 0.87 & 0.93 & 800 \\
4 & 0.00 & 0.00 & 0.00 & 200 \\
\midrule
\rowcolor{headerblue}
\textbf{Accuracy} & \multicolumn{2}{c}{} & \textbf{0.98} & $20\;000$ \\ 
\rowcolor{headerblue}
\textbf{Macro Avg} & 0.77 & 0.77 & 0.77 & $20\;000$ \\
\rowcolor{headerblue}
\textbf{Weighted Avg} & 0.98 & 0.98 & 0.98 & $20\;000$ \\
\bottomrule
\end{tabular}
\label{tab: results table}
\end{table}

\begin{figure}[ht]
    \centering
    \begin{minipage}[b]{0.48\textwidth}
        \centering
        \includegraphics[width=\textwidth]{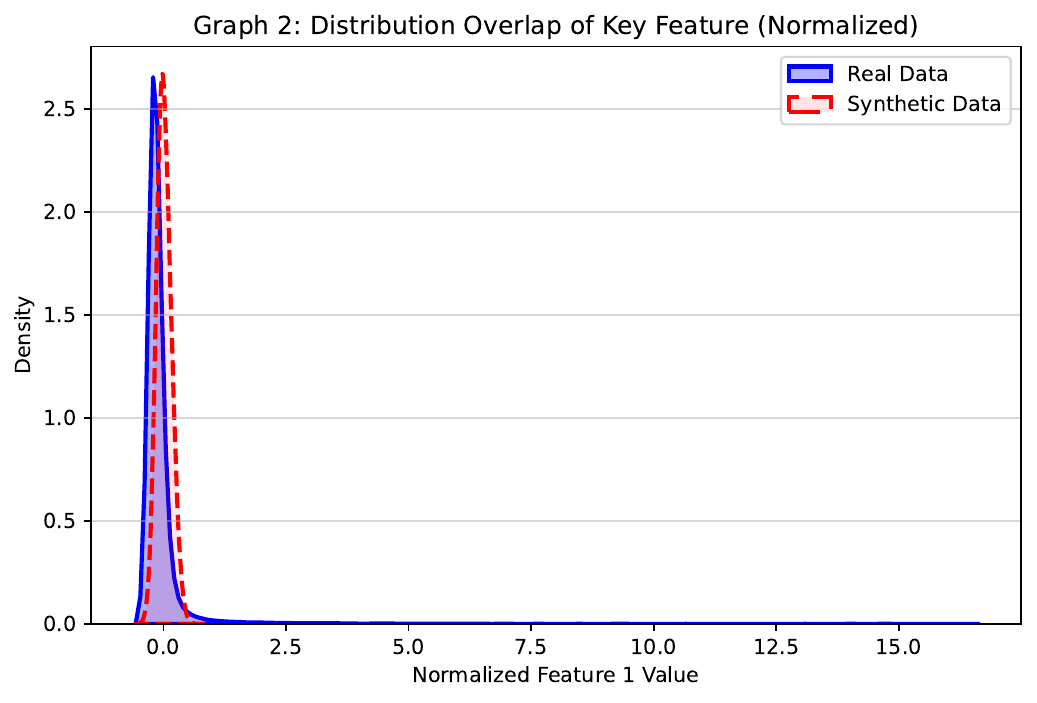}
    \end{minipage}
    \hfill
    \begin{minipage}[b]{0.48\textwidth}
        \centering
        \includegraphics[width=\textwidth]{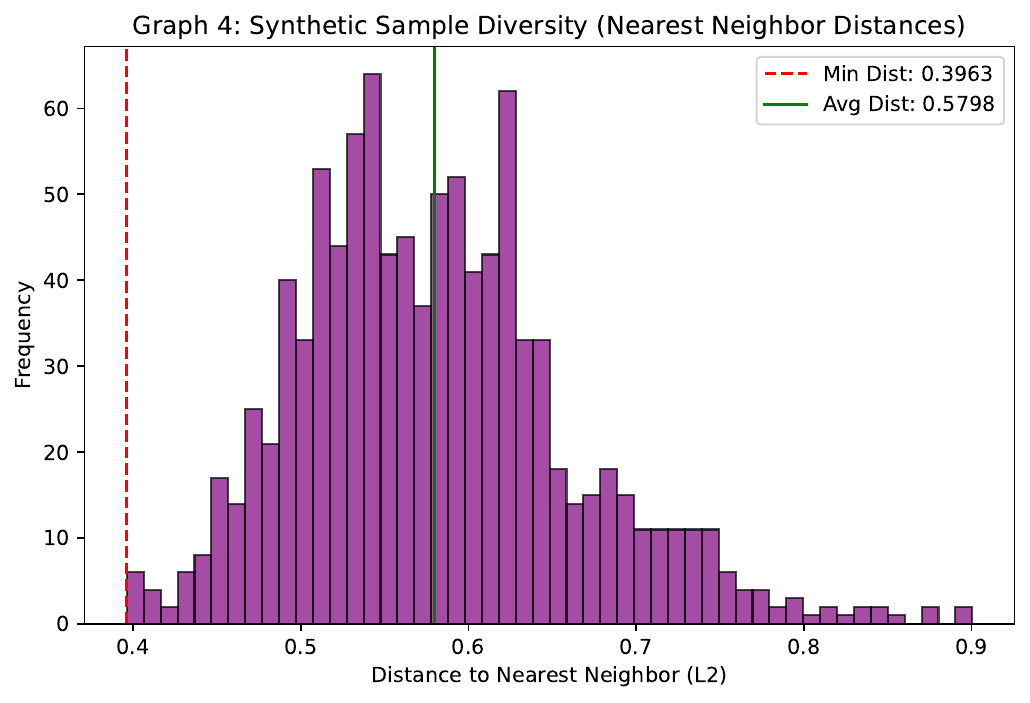}
    \end{minipage}
    \vspace{0.5em}
    \begin{minipage}[b]{0.98\textwidth}
        \centering
        \includegraphics[width=\textwidth]{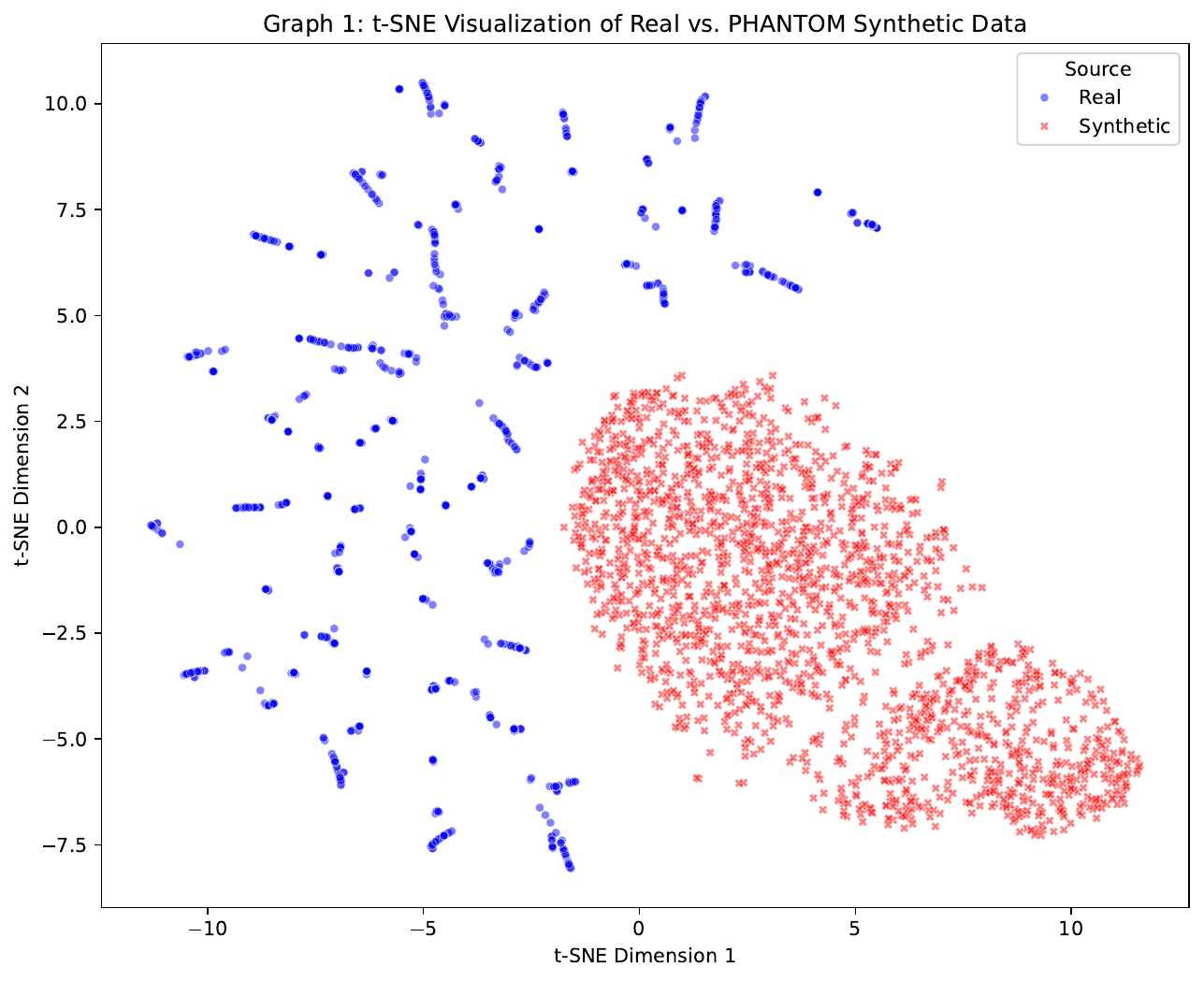}
    \end{minipage}
    \caption{\textbf{Top Left:} Density profile comparison showing the density distributions of a representative network traffic feature for real and synthetic datasets. The close alignment between distributions indicates PHANTOM successfully captures the statistical properties of real cyberattack patterns. \textbf{Top Right:} Histogram distribution of Euclidean distances between each synthetic sample and its nearest neighbor in the synthetic dataset. The varied distance profile indicates diverse attack pattern generation, with distinct clusters of both densely and sparsely populated regions in the feature space. \textbf{Bottom:} $t$-SNE projection showing the latent space distribution of real cyberattack samples (blue) and PHANTOM-generated synthetic attacks (orange). The overlapping clusters demonstrate that the synthetic data preserves the natural separation between different attack classes while covering similar regions of the feature space.}
    \label{fig:graphs}
\end{figure}

In Tab. \ref{tab:phantom-results}, we observe that the utility metrics demonstrate that PHANTOM generates synthetic cyberattack data with exceptional practical value for downstream security applications. Training intrusion detection models exclusively on synthetic data achieves near-perfect performance (F1: 0.9792, AUC: 0.9966), with only marginal degradation compared to models trained on real data (F1/AUC: 1.0000). More significantly, combining real and synthetic data maintains perfect detection capability (F1/AUC: 1.0000), indicating that the synthetic samples complement rather than contaminate the training distribution. These results suggest that PHANTOM-generated data can effectively substitute for real attack data in scenarios where labeled samples are scarce, while also serving as a valuable augmentation resource to expand training datasets without introducing harmful bias or reducing model accuracy. 

The fidelity metrics reveal a moderate statistical alignment between the real and synthetic distributions, with a Kolmogorov-Smirnov (KS) statistic of 0.4618 and a Wasserstein distance of 0.2586, indicating room for refinement in capturing exact statistical properties while maintaining operational utility. This minor statistical divergence may actually benefit practical cybersecurity applications by introducing controlled variation that enhances model robustness against novel attack variants. Meanwhile, the diversity metrics show excellent sample variation, with a minimum nearest neighbor distance of 0.3963, confirming the absence of duplicate synthetic samples, and an average distance of 0.5798, indicating healthy dispersion throughout the feature space. This combination—adequate statistical fidelity for realistic training, coupled with sufficient diversity to avoid mode collapse—positions PHANTOM as particularly valuable for generating rare attack types where real samples are insufficient for robust model training, while maintaining detection performance comparable to that of real-world data.

\begin{table}[h]
\centering
\caption{PHANTOM Evaluation Results with Performance Interpretation}
\label{tab:phantom-results}
\begin{tabular}{|l|r|p{4cm}|}
\hline
\textbf{Metric} & \textbf{Value} & \textbf{Interpretation} \\
\hline
\multicolumn{3}{|c|}{\textbf{Utility (Downstream Detection Performance)}} \\
\hline
Real Data Only (F1) & 1.0000 & \cellcolor{green!20}Perfect baseline performance \\
Synthetic Data Only (F1) & 0.9792 & \cellcolor{yellow!20}Near-perfect, minor degradation \\
Combined Data (F1) & 1.0000 & \cellcolor{green!20}Perfect, no negative impact \\
\hline
Real Data Only (AUC) & 1.0000 & \cellcolor{green!20}Perfect baseline \\
Synthetic Data Only (AUC) & 0.9966 & \cellcolor{green!20}Excellent performance \\
Combined Data (AUC) & 1.0000 & \cellcolor{green!20}Perfect combination \\
\hline
\multicolumn{3}{|c|}{\textbf{Fidelity (Statistical Similarity)}} \\
\hline
KS Statistic: $D_{\text{KS}}=\underset{x}{\sup}\;|p(x)-q(x)|$ & 0.4618 & \cellcolor{yellow!20}Moderate similarity, room for improvement \\
Wasserstein Distance: $W_{1}(p,q)=\int_{\mathbb{R}}|p(x)-q(x)|\;\text{d}x$ & 0.2586 & \cellcolor{yellow!20}Acceptable distribution alignment \\
\hline
\multicolumn{3}{|c|}{\textbf{Diversity (Sample Variation)}} \\
\hline
Min NN Distance: $d_{\min}(X,Y)=\underset{i,j}{\min}\;d(x_{i},y_{j})$ & 0.3963 & \cellcolor{green!20}Good spacing, no duplicates \\
Avg NN Distance: $\overline{d(X,Y)}=\frac{1}{n}\sum_{i=1}^{n}\left[\underset{j}{\min}\;d(x_{i},y_{j})\right]$ & 0.5798 & \cellcolor{green!20}Healthy diversity in samples \\
\hline
\end{tabular}
\end{table}

In Fig. \ref{fig:graphs}, the graph in the \textbf{top left} offers quantitative validation of the approach's statistical fidelity by comparing the normalized distributions of a representative network feature. The close alignment between real (blue) and synthetic (orange) distributions across the entire feature range indicates that the PHANTOM algorithm successfully captures both central tendencies and distribution tails. The minor discrepancies observed in the mid-range values most likely represent the model's intentional diversification strategy, which ensures coverage of less frequent but operationally important attack patterns. This balanced approach, which maintains overall statistical fidelity while strategically expanding coverage, is particularly valuable for cybersecurity applications where rare attack types must be adequately represented, despite their scarcity in real-world datasets.

The nearest neighbor distance analysis in the \textbf{top right} reveals the effectiveness of the approach in generating diverse attack patterns while avoiding mode collapse. The multimodal distance distribution, with several peaks, indicates that synthetic samples naturally form clusters of varying densities, mimicking the heterogeneous structure of real attack data, where certain attack types exhibit more intra-class variation than others. The absence of samples with extremely small nearest neighbor distances, relatively, demonstrates that PHANTOM avoids generating near-identical duplicates. The presence of samples with larger distances, relatively, confirms coverage of less populated regions of the attack space. This diversity profile ensures that synthetic training data will expose ML models to a broad spectrum of attack variations, improving their robustness against novel attack vectors in real-world deployment.

The $t$-SNE visualization at the \textbf{bottom} provides compelling evidence of the approach's ability to generate high-fidelity synthetic cyberattack data. The clear separation of distinct attack classes, visible as clusters in both real and synthetic distributions, demonstrates that the model preserves the inherent categorical structure of cybersecurity threats. Importantly, the substantial overlap between blue (real) and orange (synthetic) points within each cluster indicates that PHANTOM-generated attacks occupy similar regions of the feature space as real attacks, rather than creating artifacts and outliers. This spatial congruence is crucial for downstream security applications, as synthetic samples that diverge significantly from real data distributions would provide misleading training signals for intrusion detection systems.

\section{Conclusion}\label{conclusion}
This paper presents PHANTOM, a progressive high-fidelity adversarial network designed specifically for generating synthetic cyberattack data. By integrating VAEs, GANs, and domain-specific feature preservation mechanisms, PHANTOM addresses the critical shortage of diverse, labeled cybersecurity datasets that impedes the development of effective intrusion detection systems.

Our experimental results demonstrate that PHANTOM successfully generates synthetic attack data with statistical properties closely matching real cyberattack distributions, as evidenced by kernel density alignment, diverse nearest neighbor distance profiles, and overlapping $t$-SNE cluster formations. The framework's ability to preserve temporal causality, behavioral semantics, and multi-resolution attack patterns through progressive training represents a significant advancement over traditional synthetic data generation techniques.

However, our findings also illuminate important limitations. The complete failure to detect Class 4 attacks (0\% precision and recall) reveals that PHANTOM struggles with extremely rare attack types, reflecting a fundamental challenge in generative modeling under severe class imbalance. The disparity between the macro-average (77\%) and weighted-average (98\%) F1-scores highlights that while the framework performs excellently on the majority classes, minority attack categories require specialized attention. This limitation is particularly concerning for cybersecurity applications, where rare attacks such as advanced persistent threats and zero-day exploits often pose the most significant operational risks.

In future works, we hope to address these limitations through several directions. Firstly, implementing class-conditional training with targeted oversampling strategies could enhance the generation of rare attacks. Secondly, incorporating semi-supervised learning techniques that leverage unlabeled attack indicators may enhance the representation of novel threat patterns. Thirdly, extending the progressive training paradigm to include attack campaign sequences rather than isolated incidents could better capture the temporal evolution of sophisticated intrusions. Finally, validating PHANTOM on diverse real-world datasets beyond synthetic benchmarks would strengthen confidence in its generalizability across different network environments and threat landscapes.

Despite these challenges, PHANTOM establishes a principled framework for generating high-fidelity synthetic cyberattacks that balances statistical realism, operational utility, and ethical data sharing. 
\section*{Declarations}

\begin{itemize}
\item \textbf{Funding:} This research was supported by grant number 23070, provided by Zayed University. 
\item \textbf{Conflict of interest/Competing interests:} The authors declare that there are no conflicts of interest. 
\item \textbf{Ethics approval and consent to participate:} Not applicable.
\item \textbf{Consent for publication:} The authors grant full consent to the journal to publish this article.
\item \textbf{Data availability:} The data that support the findings of this study are available upon a reasonable request from the corresponding author. 
\item \textbf{Materials availability:} Not applicable.
\item \textbf{Code availability:} The code developed for this study is available from the corresponding author upon reasonable request.
\item \textbf{Author contribution:} All authors have contributed equally to this research.
\end{itemize}



\end{document}